\documentclass[doublecol]{epl2_format} 
\pdfoutput=1

\usepackage{graphicx}
\usepackage{amssymb}
\usepackage{amsmath}

\title{Effective surface interactions mediated by adhesive particles}

\author{Bartosz R\'{o}\.{z}ycki, Reinhard Lipowsky and Thomas R. Weikl}

\institute{Max Planck Institute of Colloids and Interfaces, Department of Theory and Bio-Systems, 14424 Potsdam, Germany}


\abstract{
In biomimetic and biological systems, interactions between surfaces are often mediated by adhesive molecules, nanoparticles, or colloids dispersed in the surrounding solution. We present here a general, statistical-mechanical model for two surfaces that interact via adhesive particles. The effective, particle-mediated interaction potential of the surfaces is obtained by integrating over the particles' degrees of freedom in the partition function. Interestingly, the effective adhesion energy of the surfaces exhibits a maximum at intermediate particle concentrations, and is considerably smaller both at low and high concentrations. The effective adhesion energy corresponds to a minimum in the interaction potential at surface separations slightly larger than the particle diameter, while a secondary minimum at surface contact reflects depletion interactions. Our results can be generalized to surfaces with specific receptors for solute particles, and have direct implications for the adhesion of biomembranes and for phase transitions in colloidal systems.
}

\begin{document}

\maketitle

\section{Introduction}

The adjustment of surface interactions is crucial for controlling the phase behavior of colloidal systems \cite{Anderson02} and the adhesiveness of biological cells and membranes \cite{Alberts02}. These interactions are often dominated by the composition of the surfaces, which may be charged, hydrophobic, etc. In some systems, the interactions are also strongly affected by molecules or particles in the surrounding medium. The concentration of these particles is an additional control parameter for the surface interactions, a parameter that is often easier to adjust than the surface composition, and can be varied over a wider range than external parameters such as temperature. 

On the one hand, non-adhesive particles can induce attractive `depletion' interactions between surfaces, because close contact of the surfaces reduces the excluded volume for the particles \cite{Asakura54,Dinsmore96}. On the other hand, adhesive particles can directly bind two surfaces together. For example, the surface interactions and colloidal phase of membrane-coated silica beads \cite{Baksh04,Winter06} and the force between membrane-coated mica surfaces \cite{Hu04} have been altered by adding soluble, adhesive proteins. Multivalent ions such as chromium can induce the adhesion of lipid membranes, presumably by crosslinking the polar headgroups of lipids in apposing membranes \cite{Franke06}. Linker proteins that interconnect membrane receptors are known to assist biomembrane adhesion \cite{Alberts02}, which has also been utilized in biomimetic experiments \cite{Saterbak96,Orsello01}.

\begin{figure}
\begin{center}
\resizebox{\columnwidth}{!}{\includegraphics{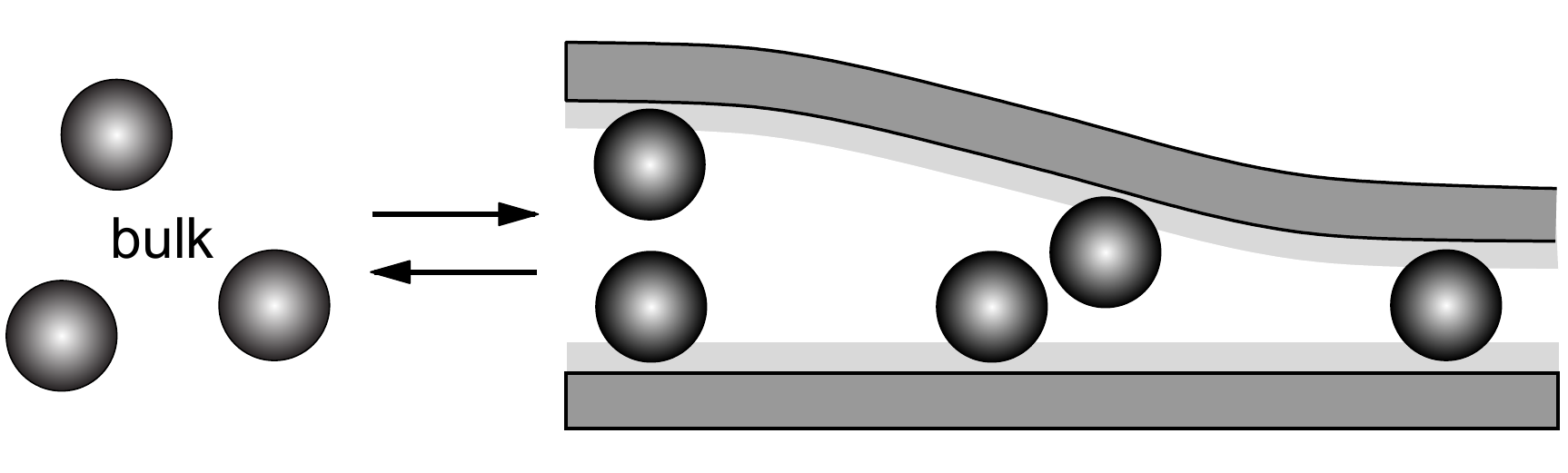}}
\caption{Two surfaces in contact with a solution of adhesive particles. A particle can bind the two surfaces together for surface separations slightly larger than the particle diameter (particle on the right). At large separations,  the particles can only bind to one of the surfaces (particles on the left). }
\label{cartoon}
\end{center}
\end{figure}
\begin{figure*}
\begin{center}
\resizebox{2\columnwidth}{!}{\includegraphics{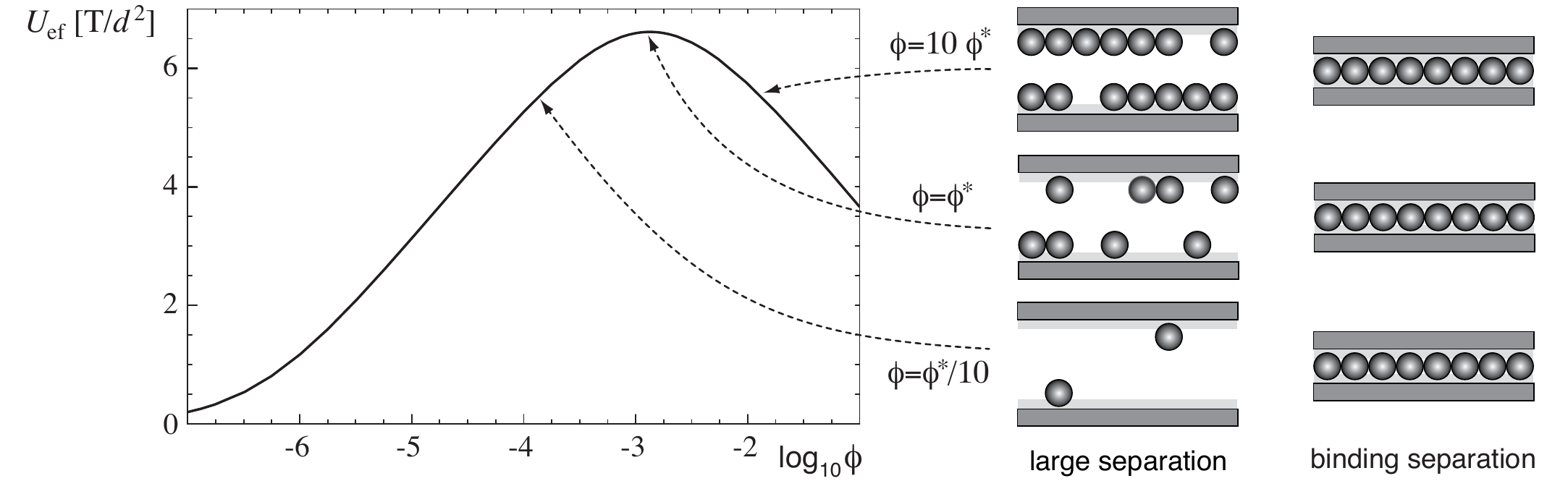}}
\caption{Effective adhesion energy $U_{\rm ef}$ of the surfaces, given in eq.~(\ref{gamma2}), as a function of the particle bulk volume fraction $\phi$ for the binding energy $U= 8T$ and $q = 0.25$. The effective adhesion energy is maximal at the optimal bulk volume fraction $\phi^{\star} \approx e^{-U/T}/ q \approx 1.34 \cdot 10^{-3}$. At the optimal volume fraction, the particle coverage of two planar parallel surfaces is 50\% for large separations, and almost 100\% for small separations at which particles can bind to both surfaces. The surface coverage at these small separations remains close to 100\% for volume fractions $\phi^{\star} /10 \lesssim \phi \lesssim 10 \, \phi^{\star}$, while the coverage $c_{\infty} = \phi / ( \phi + \phi^{\star} )$ for large separations changes between approximately 9\% and 90\% in this example. }
\label{g}
\end{center}
\end{figure*}

Adhesive particles can bind two surfaces together if the separation of the surfaces is equal to or slightly larger than the particle diameter, see fig.~\ref{cartoon}. At larger separations, the particles can only bind to one of the surfaces. In this letter, we consider particles that exhibit short-ranged, attractive interactions with the surfaces, and repulsive hard-sphere interactions with each other. Our central result is that the effective, particle-mediated adhesion energy of the surfaces is given by
\begin{equation}
U_{\rm ef} \approx \frac{T}{d^2} \ln \frac{ 1+ q \,\phi \, e^{2U/T} }{ \left( 1+ q \,\phi \, e^{U/T} \right)^2 } 
\label{gamma2}
\end{equation}
with a dimensionless coefficient $q$ for small bulk volume fractions $\phi\ll 1$ of the particles and large binding energies $U$ with $e^{U/T}\gg 1$. Here, $d$ is the particle diameter, and $T$ denotes the temperature in energy units. 

Interestingly, the effective adhesion energy (\ref{gamma2}) of the surfaces exhibits a maximum at an optimum bulk volume fraction of the particles, see fig.~\ref{g}. At this volume fraction, the particle coverage of two planar parallel surfaces turns out to be close to 50\% for large separations (`half coverage'), and 100\% (`full coverage') for short, binding separations, see fig.~\ref{g}. Bringing the surfaces from large separations within binding separations thus does not `require' desorption or adsorption of particles at the optimum volume fraction. The existence of an optimum particle volume fraction has interesting, experimentally testable implications, for example `re-entrant transitions' in which surfaces or colloidal objects first bind with increasing concentration of adhesive particles, and unbind again when the concentration is further increased  beyond the optimum concentration. 
 
Even though our derivations of eq.~(\ref{gamma2}) are based on some simplifying assumptions, the effective adhesion energy (\ref{gamma2}) should be applicable in general since it can simply be understood as a difference of two Langmuir adsorption free energies per binding site: (i) the adsorption free energy $(T/d^2)\ln \left( 1+ q\, \phi \, e^{2U/T}\right)$ for small surface separations at which a particle binds both surfaces with total binding energy $2U$, and (ii) the adsorption free energy $(T/d^2)\ln \left(1+ q \, \phi \, e^{U/T}\right)$ for large surface separations, counted twice in (\ref{gamma2}) because we have two surfaces. These Langmuir adsorption free energies  result from a simple two-state model in which a particle is either absent (Boltzmann weight $1 - \phi\approx 1$) or present (Boltzmann weights $q\, \phi \, e^{2U/T}$ and $q\, \phi \, e^{U/T}$, respectively) at a given binding site, see e.g.\ \cite{davis96}. The factor $q$ depends on the degrees of freedom of a single adsorbed particle and has to be determined from more detailed binding models. 

In the following, we will study particles of diameter $d$ that are attracted to the two surfaces within the binding range $r$. We will first consider a 3-dimensional lattice gas model with $d = r$ that leads to relation (1) with $q = 1$. We will then study a more general  tube model with arbitrary values of $d$ and $r$ from which we obtain another derivation of (1) with $q = r/d$. Finally, we generalize our main result (\ref{gamma2}) to surfaces with specific receptors for solute particles or molecules. 

\section{Lattice gas model}

The simplest model for a gas of adhesive particles between two parallel and planar surfaces is obtained by discretizing the space between the surfaces into a cubic lattice. In this model, the hard-core interactions between the particles are incorporated by choosing a lattice spacing $d$ equal to the particle diameter. Each lattice site then can contain only one particle, and the bulk volume fraction of particles is $\phi = e^{\mu/T}/(1+e^{\mu/T})$ where $\mu$ is the chemical potential. If a particle is located at a lattice site adjacent to one of the two surfaces, it gains the binding energy $U$. The separation of the surfaces is $\ell = n_\perp d$ where $n_\perp$ is the number of lattice layers between the surfaces. The total number of lattice sites between the two surfaces is $n_\parallel n_\perp$.

The free energy $F$ of this lattice gas of particles between the surfaces can be decomposed into a bulk and a surface term. The bulk free energy $f_b$, which is equal to the free energy per lattice site in the limit of large $n_\parallel n_\perp$, has the simple form $f_b = -T \ln(1 + e^{\mu/T}) = T \ln(1 - \phi)$. The surface free energy $f_s= (F-n_\parallel n_\perp f_b)/n_\parallel$ is the excess free energy per pair of apposing surface sites in the limit of large $n_\parallel$. The surface free energy depends on the separation of the surfaces, i.e.~on the number of lattice layers $n_\perp$. 
The surface free energy is given by
\begin{equation}
f_{s,n_\perp} = f_{s,\infty} = -2 \,T \ln\frac{1+e^{(U+\mu)/T}}{1+e^{\mu/T}}
\label{fsn}
\end{equation}
for $n_\perp \ge 2$ and by
\begin{equation}
f_{s,1} = -T \ln \frac{1+e^{(2 U+\mu)/T}}{1+e^{\mu/T}}
\label{fs1}
\end{equation}
for $n_\perp = 1$. The effective adhesion potential of the surfaces, defined as $V =  f_{s,n_\perp} /d^2$, thus is constant for $n_\perp \ge 2$ and has an attractive well of depth  $U_\text{ef} = (f_{s,\infty} - f_{s,1})/d^2$ at $n_\perp = 1$. After replacing the chemical potential $\mu$ in eqs.~(\ref{fsn}) and (\ref{fs1}) by the bulk volume fraction $\phi = e^{\mu/T}/(1+e^{\mu/T})$, we obtain  
\begin{equation}
U_\text{ef} = \frac{T}{d^2} \ln \frac{1- \phi + \phi \, e^{2 U/T}}{(1- \phi + \phi\, e^{U/T})^2} 
\label{Ueflattice}
\end{equation}
For small volume fractions $\phi\ll 1$ and large binding energies $U$ with $e^{U/T}\gg 1$, the effective adhesion energy (\ref{Ueflattice}) is identical with (\ref{gamma2}) for $q = 1$.

\section{Tube model}

We now consider a semi-continuous model to obtain a realistic estimate of the factor $q$ in eq.~(\ref{gamma2}) from the separation-dependent effective interaction potential of the surfaces. In this model, we discretize the space between the surfaces into tubes of the same diameter $d$ as the particles. We assume that the two apposing surfaces are on average parallel and nearly planar, with local curvature radii much larger than the diameter $d$ of the adhesive particles. The tubes are oriented perpendicular to both surfaces and contain the particles. The length $\ell$ of the tubes thus corresponds to the local separation of the surfaces. The particles can exchange between the tubes and with the bulk solution (see fig.~\ref{cartoon}). We assume that the attractive interaction of a particle with the surfaces is short-ranged and model this interaction by a square-well potential with binding energy $U$ and range $r<d/2$. A particle is thus bound to a surface with binding energy $U$ if the separation between the surface and the center of the particle is smaller than $r+d/2$. The pair interactions between particles are purely repulsive hard-sphere interactions.
For simplicity, we assume that the tubes are arranged on a square lattice. Each tube then occupies a volume $d^2 \ell$. The approximation implied by this discretization should be valid for the experimentally relevant small particle bulk volume fractions $\phi$ considered here. Since the particle-surface interaction is short-ranged, the gas of particles between the surfaces is as dilute as in the bulk for large surface separations $\ell\gg d$, except for the single adsorption layers of particles at the surfaces. For large binding energies $U$, these adsorption layers will fully cover the surfaces.

In this model, the problem of determining the effective adhesion potential of the surfaces is reduced to calculating the partition function of a one-dimensional gas of particles in a tube. Since the number of particles in a tube varies, the suitable statistical ensemble is the grand-canonical ensemble in which the chemical potential $\mu$ of the particles is fixed. The free energy is 
\begin{equation}
F = - T \ln \left[ 1+ \sum_{n=1}^{ \lfloor \ell /d \rfloor } e^{n \mu /T} \mathcal{Z}_n \right] \label{free_energy}
\end{equation}
where $\mathcal{Z}_n$ denotes the canonical partition function for a system of $n$ particles confined in the tube. The upper limit of the sum, $\lfloor \ell /d \rfloor$, equals the largest number of hard spheres of diameter $d$ that can be placed in the tube of length $\ell$. The effective adhesion potential of the surfaces is $V= (F  - f_b d^2 \ell)/d^2$, where $f_b = \lim_{\ell \to \infty} F / (d^2 \ell)$ denotes the free energy density in the bulk. The effective adhesion potential thus is again defined as the surface contribution to the free energy $F$. With eq.~(\ref{free_energy}), we obtain
\begin{equation}
V = - \frac{T}{d^2} \ln \left[ \left( 1+ \sum_{n=1}^{\lfloor \ell /d \rfloor} \mathcal{Z}_n e^{n \mu /T} \right) \exp \left( \frac{ f_b d^2 \ell}{T} \right) \right] 
\label{VV}
\end{equation}

\begin{figure}[t]
\begin{center}
\resizebox{0.99\columnwidth}{!}{\includegraphics{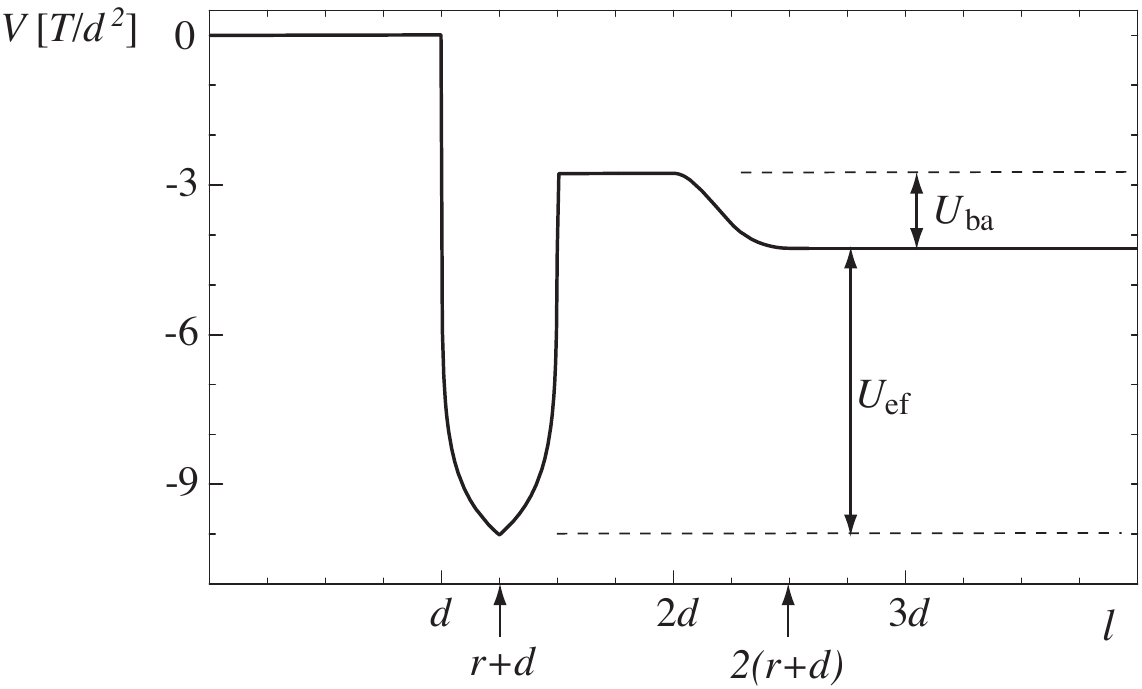}}
\caption{Effective adhesion potential $V$ as a function of the surface separation $\ell$ for the bulk volume fraction $\phi =0.01$, binding energy $U=8T$, and binding range $r =d/4$ of the particles with diameter $d$. The potential has a minimum at the surface separation $\ell =r+d$ and attains a constant value for separations $\ell >2(r+d)$. The effective adhesion energy of the surfaces, $U_{\rm ef}$, is the difference between the asymptotic and minimum value of potential $V( \ell )$. For large binding energies $U$ with $e^{U/T} \gg 1$, the effective potential has a barrier of height $U_{\rm ba}$ at surface separations $d+2r< \ell < 2d$. }
\label{Vef1}
\end{center}
\end{figure}
\begin{figure}[t]
\begin{center}
\resizebox{0.99\columnwidth}{!}{\includegraphics{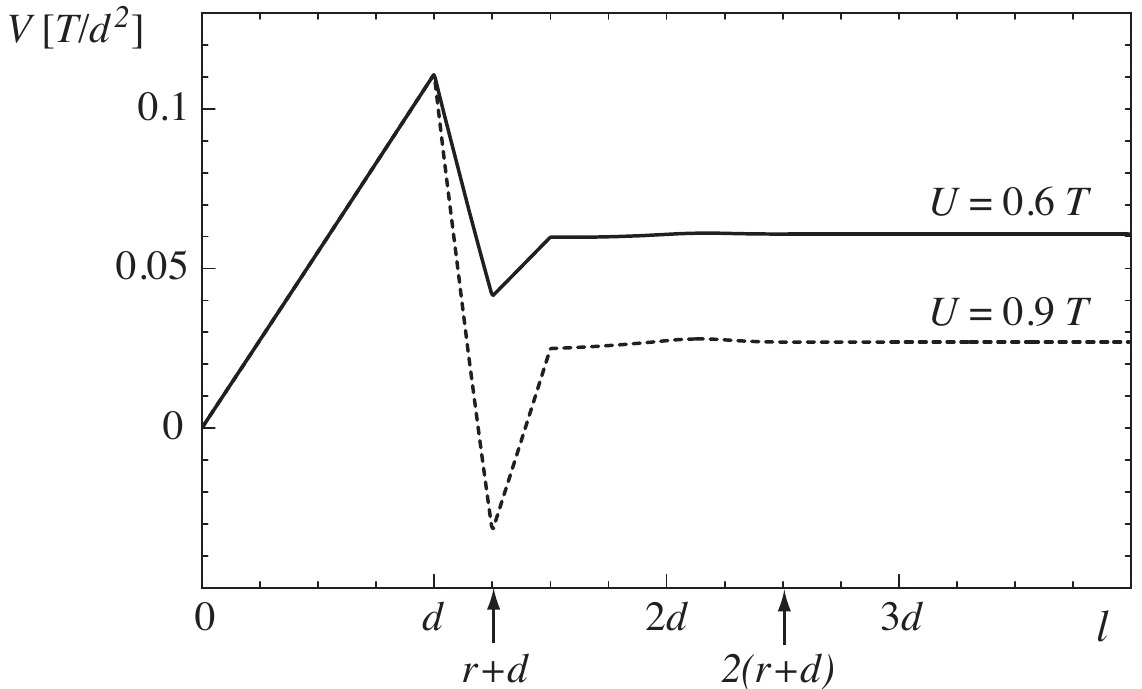}}
\caption{Effective adhesion potential $V$ as a function of the surface separation $\ell$ for the particle bulk volume fraction $\phi =0.1$,  binding range $r =d/4$, and binding energies $U= 0.9 \, T$ (lower curve) and $U=0.6 \, T$ (upper curve). The global minimum of potential $V( \ell )$ is located at surface contact $\ell =0$ for small energies $U\lesssim \frac{1}{2} T \ln(1+d/r)$ (upper curve), and at the separation $\ell =d+r$ for binding energies $U\gtrsim \frac{1}{2} T \ln(1+d/r)$ (lower curve). }
\label{Vef2}
\end{center}
\end{figure}

One-dimensional models for hard spheres have been studied extensively \cite{tonks36, lieb66, davis90, davis96, iwamatsu03, chou03}. Two well-known results that we will use in the following are: (i) The hard-sphere gas fugacity is $e^{\mu /T} \approx \left( \phi + 2 \phi^2 \right) \Lambda /d$, up to second order terms in the bulk volume fraction $\phi$ \cite{davis90}. Here, $\Lambda$ denotes the thermal de Broglie wavelength. (ii) The bulk free energy density is $f_b = - T \phi /(d^3(1 - \phi))$,  with the denominator describing the volume accessible to a sphere \cite{davis96}. In contrast to previous studies, we are interested here in the effective adhesion potential V given by eq.~(\ref{VV}). With the relations (i) and (ii) above, a virial expansion of the potential $V$ leads  to
\begin{equation}
V \approx - \frac{T}{d^2} \ln \left[ 1+ a_1 ( \ell ) \phi + a_2 ( \ell ) \phi^2 \right]
\label{V}
\end{equation}
for $\phi\ll 1$ with the expansion coefficients
\begin{equation}
\begin{array}{l}
a_1 = \left[ \Lambda \mathcal{Z}_1 (\ell) - \ell \right] /d \quad {\rm and } \\
a_2 = \left[ \left( 2d - \ell \right) \Lambda \mathcal{Z}_1 (\ell) + \Lambda^2 \mathcal{Z}_2 (\ell) - \ell d + \ell^2 /2 \right] /d^2.
\end{array}
\label{virial}
\end{equation}

The partition function $\mathcal{Z}_1$ is given by
\begin{equation}
\mathcal{Z}_1 = \frac{1}{\Lambda} \int_{d/2}^{\ell - d/2} {\rm d} x_1 \, e^{-H(x_1) /T} 
\label{Z1}
\end{equation}
with the one-particle configuration energy $H(x_1) = -U \Theta \left( d/2 +r -x_1 \right) -U \Theta \left( x_1 - \ell + d/2 +r \right)$. Here, $x_1$ denotes the center-of-mass position of the particle in the tube, and $\Theta$ is the Heaviside step function with $\Theta (x) =1$ for $x>0$ and $\Theta (x) =0$ for $x<0$. We have assumed here that the tube length $\ell$ is larger than the particle diameter $d$. For tube lengths $\ell > 2d$, the partition function $\mathcal{Z}_2$ is
\begin{equation}
\mathcal{Z}_2 = \frac{1}{\Lambda^2} \int_{3d/2}^{\ell - d/2} {\rm d} x_2 \int_{d/2}^{x_2 -d} {\rm d} x_1\, e^{-H( x_1 , x_2 ) /T} 
\label{Z2}
\end{equation}
with the two-particle configuration energy $H(x_1, x_2) = -U \Theta \left( d/2 +r -x_1 \right) -U \Theta \left( x_2 - \ell + d/2 +r \right)$. For tube lengths $\ell < d$ and  $\ell < 2 d$, respectively, the partition functions $\mathcal{Z}_1$ and $\mathcal{Z}_2$ are zero. 

To perform the integrations in eqs.~(\ref{Z1}) and (\ref{Z2}), it is helpful to rewrite the integrand by making use of the simple relation $e^{U \Theta (x) /T} = 1+ (e^{U/T} -1) \Theta (x)$. The integrations then lead to
\begin{eqnarray}
\mathcal{Z}_n = 
\frac{e^{2U/T}}{\Lambda^n} \bigg[
\left( 1- e^{-U/T} \right)^2 {\rm f}_n \left( \ell  -2 r \right) \nonumber \\
- 2 \left( 1- e^{-U/T} \right) {\rm f}_n \left( \ell - r \right)
+ {\rm f}_n \left( \ell \right) \bigg]
\label{Zn}
\end{eqnarray}
with the auxiliary function f$_n (l)=(l-nd)^n \Theta (l-nd) /n!$. The effective adhesion potential $V( \ell )$ finally is obtained from inserting this result into eqs.~(\ref{V}) and (\ref{virial}). 

For small bulk volume fractions $\phi$ considered here, the effective adhesion potential is constant for separations $\ell>2(d+r)$, with the asymptotic value
\begin{eqnarray}
V_{\infty} \approx - \frac{T}{d^2} \ln \bigg[ 1+ \phi \left( 2 \frac{r}{d} \left( e^{U/T} -1 \right) -1 \right) \nonumber \\
+ \phi^2 \frac{r^2}{d^2} \left( e^{U/T} -1 \right) \left( e^{U/T} -2 \right) \bigg] .
\label{Vinfty}
\end{eqnarray}
The adhesion potential has a local minimum 
\begin{equation}
V_{\rm min} \approx-\frac{T}{d^2} \ln \left[ 1+ \phi \left( \frac{r}{d} \left( e^{2U/T} -1 \right) -1 \right) \right] 
\label{V_d_plus_r}
\end{equation}
at the surface separation $\ell = r+d$, see figs.~\ref{Vef1} and \ref{Vef2}. At this separation, terms of order $\phi^2$ are negligible since the two-particle partition function $\mathcal{Z}_2$ is 0. The virial expansion coefficients $a_1$ and $a_2$ then are of the same magnitude, and $a_2 \phi^2$ is much smaller than $a_1\phi$ for $\phi\ll 1$. 
 
Depletion interactions are reflected in a second minimum of the effective potential at surface contact $\ell =0$, see fig.~\ref{Vef2}. For surface separations $0 < \ell <d$, the potential $V( \ell ) = \ell T \phi / [(1- \phi)d^3]$ increases linearly with $\ell$ due to the depletion forces. By definition, the effective potential vanishes at surface contact $\ell =0$. The two minima thus have equal depths for $V_{\rm min} = 0$, i.e.~at the binding energy $U_0 \approx \frac{1}{2} T \ln(1+d/r)$. For binding energies $U>U_0$, the global minimum of the potential $V( \ell )$ is located at $\ell=r+d$.  
The effective adhesion energy of the surfaces then is $U_{\rm ef} = V_{\infty}  - V_{\rm min}$ with $V_{\infty}$  and $V_{\rm min}$ as in eqs. (\ref{Vinfty}) and (\ref{V_d_plus_r}). For large binding energies $U$ with $e^{U/T} \gg 1$, we obtain the effective adhesion energy (\ref{gamma2}) with $q = r/d$.

The adhesion energy $U_{\rm ef}$ is maximal at the bulk volume fraction $\phi = \phi^{\star}$ with $\phi^{\star} \approx e^{-U/T} \, d/r$, see fig.~\ref{g}. At this optimal volume fraction, the coverage of the unbound surfaces, $c_{\infty} = -(d^2 /2) (\partial V_{\infty} / \partial U) \approx \phi /( \phi + \phi^{\star})$, is 50\%. In contrast, the coverage of the bound surfaces at separation $\ell =d+r$ is $c_{\rm min} = - (d^2 /2) (\partial V_{\rm min} / \partial U) \approx \phi /( \phi + \phi^{\star} e^{-U/T})$ and approaches 100\% for $\phi \approx \phi^{\star}$.

Besides the two minima at $\ell =0$ and $\ell = d+r$, the effective adhesion potential has a barrier at surface separations $d+2r< \ell < 2(d+r)$ for large binding energies, see fig.~\ref{Vef1}. At these separations, only a single particle fits between the surfaces, but this particle can just bind one of the surfaces. The particle thus `blocks' the binding site at the apposing surface, see particles in the center of fig.~\ref{cartoon}.
At the surface separation $\ell =2d$ within the potential barrier, the effective adhesion potential $V( \ell )$ attains the value
\begin{equation}
V_{\rm ba} \approx - \frac{T}{d^2} \ln \left[ 1+ \phi \left( 2 \frac{r}{d} \left( e^{U/T} -1 \right) -1 \right) \right] 
\end{equation}
for small bulk volume fractions $\phi$. At large binding energies $U$ with $e^{U/T} \gg 1$, the barrier height $U_{\rm ba} =  V_{\infty}  - V_{\rm ba}$ then is  
\begin{equation}
U_{\rm ba} \approx \frac{T}{d^2} \ln \frac{\left( 1+ \phi \, e^{U/T} \, r/d \right)^2}{1+ 2 \phi \, e^{U/T} \, r/d } 
\label{U_ba}
\end{equation}
\begin{figure}
\begin{center}
\resizebox{\columnwidth}{!}{\includegraphics{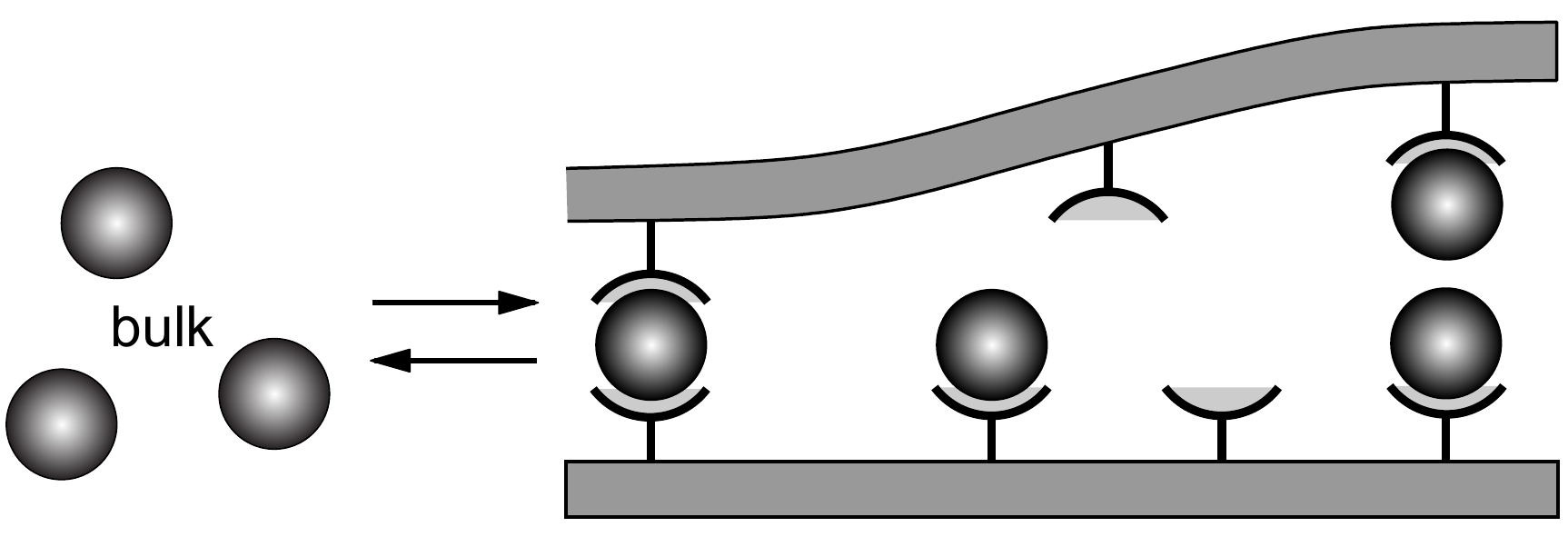}}
\caption{Two surfaces, e.g. lipid membranes, with specific receptors for solute particles or molecules in the bulk. For small surface separations, a particle can link the surfaces together by binding to two receptors at apposing surface sites (particle on the left).}
\label{cartoon_two}
\end{center}
\end{figure}
%

\section{Surfaces with specific receptors}

The effective adhesion energy (\ref{gamma2}) can be generalized to cases in which the adhesive particles specifically bind to receptor sites or molecules at the surfaces, e.g.~to receptors  in fluid lipid membranes. An example are biotinylated lipids, which can be crosslinked by streptavidin \cite{Albersdoerfer97,Leckband94}. In principle, the two membranes can contain the same type of receptors, as in fig.~\ref{cartoon_two}, or different types of receptors. We characterize the receptors in membrane 1 by their binding energy $U_1$ and chemical potential $\nu_1$, and the receptors in membrane 2 by the  binding energy $U_2$ and chemical potential $\nu_2$. For simplicity, we assume the same binding range $r$ for both types of receptors. The chemical potential of the receptors is the free energy difference between a membrane patch of size $d^2$ containing a receptor molecule, and a membrane patch of the same size without receptor \cite{Weikl06}. In the absence of adhesive particles, the chemical potentials $\nu_1$ and $\nu_2$ are directly related to the area fractions $\alpha_1 = e^{\nu_1/T}/(1+e^{\nu_1 /T})$ and $\alpha_2 = e^{\nu_2 /T}/(1+e^{\nu_2 /T})$ of the receptors in the two membranes.

Let us first consider the Langmuir adsorption free energy of a single membrane with receptors. Since each patch of the membrane can attain four possible states, the Langmuir free energy per patch area $d^2$ is $(T/d^2) \ln \left(1+ e^{\nu_i /T} + q \, \phi + q \, \phi \, e^{(U_i + \nu_i )/T} \right)$ with $i=1$ or $2$. Here, $e^{\nu_i /T}$ is the Boltzmann weight for the state in which a receptor is present in the patch but no particle is within binding range, $q \phi$ is the Boltzmann weight for having a particle within binding range in absence of a receptor, and $q \phi \, e^{(U_i + \nu_i )/T}$ is the Boltzmann weight for a receptor bound to an adhesive particle. The prefactor $q$ depends on the degrees of freedom of bound particles. In our model, we obtain $q=r/d$, see previous section. 

At large separations, the Langmuir adsorption free energy of two membranes simply is the sum of the adsorption energies above. At the optimum binding separation, however, the Langmuir adsorption energy of two apposing membrane patches is $(T/d^2) \ln [ (1+e^{\nu_1/T})(1+e^{\nu_2 /T})+ q \phi (1+ e^{(U_1 + \nu_1)/T})(1+ e^{(U_2 +\nu_2)/T}) ]$. As before, the effective particle-mediated adhesion energy of the membranes is  the difference between the Langmuir adsorption energies at large separation and the optimum binding separation. In the symmetric case with binding energies $U= U_1 =U_2$ and chemical potentials $\nu = \nu_1 = \nu_2$, we obtain
\begin{equation}
U_{\rm ef} \approx \frac{T}{d^2} \ln \frac{ 1+ \phi \left( 1 - \alpha
+\alpha e^{U/T} \right)^2 \, r/d }{ \left( 1+ \phi \left( 1- \alpha +
\alpha e^{U/T} \right) r/d \right)^2 }  
\label{gamma3}
\end{equation}
with $\alpha = e^{\nu /T}/(1+e^{\nu /T})$. The effective adhesion energy (\ref{gamma3}) can be obtained directly from eq.~(\ref{gamma2}) by replacing the Boltzmann factor $e^{U/T}$ with the term $1 - \alpha + \alpha e^{U /T}$. For $\alpha =1$ where each membrane patch contains a receptor, the eqs.~(\ref{gamma2}) and (\ref{gamma3}) become identical. The effective adhesion energy (\ref{gamma3}) is maximal at the bulk volume fraction $\phi^{\star} \approx (d/r)/(1 - \alpha + \alpha e^{U/T})$.

\section{Discussion and outlook}

We have determined the effective interactions between two surfaces in contact with adhesive particles. Our results apply to a wide range of surfaces, which may be soft or rigid, and planar or non-planar. For non-planar surfaces, the local surface separation $\ell$ varies, and the total particle-mediated interaction energy is obtained by integration over the separation-dependent local interaction $V(\ell)$ defined in (\ref{VV}). The overall interaction potential of the surfaces is a superposition of the particle-mediated interactions with direct interactions, such as van der Waals, electrostatic and hydration forces. For large particle radii $d$, the particle-mediated interactions should dominate since the particle-bound surfaces have a separation close to $d+r$, where $r$ is the range of the attractive surface-particle interaction. The interactions induced by the particles then can be measured directly, e.g.~via the surface-force apparatus \cite{Hu04}, or inferred from the phase behavior of colloidal systems \cite{Baksh04,Winter06}.  

The effective adhesion energy (\ref{gamma2}) has been obtained from two different models: (i) from a 3-dimensional lattice gas model; and (ii) from a more elaborate tube model for hard spheres. In both cases, we discretized the space in order to incorporate the hard-core interaction between the particles in an analytically tractable manner. In the lattice gas,  we discretized all three coordinates of the particles. In the tube model, the coordinate perpendicular to the two surfaces was taken to be continuous. As far as relation (\ref{gamma2}) is  concerned, the only difference between the  lattice gas model and the tube model is that they give somewhat different expressions for the dimensionless coefficient $q$. Thus, it is rather plausible that the effective attractive interaction (\ref{gamma2}) also applies to hard spheres with  three continuous spatial coordinates. This  proposition can be checked by Monte Carlo or Molecular Dynamics simulations, which have been previously used to study hard spheres confined between nonadhesive surfaces \cite{Chu94,Schmidt97,Schoen98,Fortini06,Mittal08}. 

We have neglected that flexible surfaces such as lipid  membranes can wrap around adhesive particles \cite{Lipowsky98,Deserno03,Bickel03,Fleck04}. A partial wrapping  leads to effective, surface-mediated interactions between the adsorbed particles \cite{Weikl03,Mueller05and07,Reynwar07}, and to cooperativity effects in adsorption \cite{Hinderliter06}. We have also assumed that the receptors considered in the last section are rigid molecules and not flexible tethers \cite{Jeppesen01,Moreira03,Martin06}, which seems to be a good assumption for most biological receptors. In addition, we have neglected direct, long-ranged interactions of the adhesive particles, e.g.~electrostatic repulsion of charged particles. At small bulk concentrations of the particles, repulsive interactions will mainly affect the packing density of the adsorbed particles and, thus, the concentration of available binding sites. 
For charged solutes, such as multivalent ions \cite{Franke06} or charged proteins \cite{May00} adhering to lipid membranes, the average particle separation at maximum surface coverage is affected by the salt-dependent screening length of the solutes at the surfaces. 

Thermal shape fluctuations of lipid membranes lead to an unbinding transition if the fluctuation-induced repulsion exceeds  the effective adhesion energy $U_{\rm ef}$. The character of the transition depends on the barrier in the effective adhesion potential. According to scaling arguments \cite{Lipowsky94}, the unbinding transition of the membranes is discontinuous for strong barriers with $U_{\rm ba} l_{\rm ba}^2 > c \, T^2/\kappa$  where $c$ is a dimensionless coefficient of order $0.01$ \cite{Weikl02}, and continuous for weak potential barriers. The thickness of the barrier here is $l_{\rm ba} \approx d$, see fig.~\ref{Vef1}, and $\kappa$ is the bending rigidity of the lipid membranes.  At the optimum bulk volume fraction $\phi^{\star} \approx  e^{-U/T} \, d/r$, the effective barrier strength is $U_{\rm ba}d^2  \approx 0.3 \, T$ and thus clearly beyond the threshold for discontinuous unbinding, since the bending rigidity of lipid membranes typically is between 10 and 20~$T$.

In this letter, we have considered equilibrium aspects of adhesion. The unbinding dynamics of surfaces with multiple receptor-ligand bonds under a pulling force has been studied in \cite{Seifert00and02,Prechtel02,Erdmann04}. If receptors in apposing surfaces are connected via adhesive solutes, each bond between the surfaces consists of two molecular bonds in series. Such serial bonds can break at either of their bonds, and therefore have been found to break earlier than single receptor-ligand bonds under an applied force \cite{Saterbak96,Orsello01}.

\end{document}